\documentclass[letterpaper,aps,prd,twocolumn,showpacs,preprintnumbers,nofootinbib]{revtex4}
\usepackage{graphicx}
\usepackage[intlimits,centertags]{amsmath}
\usepackage{amssymb,amsfonts}
\usepackage{gensymb}
\usepackage{mathrsfs}
\usepackage{hyperref}

\begin{document}

\title{Minimal Cosmogenic Neutrinos}

\author{Markus~Ahlers} 
\affiliation{Wisconsin IceCube Particle Astrophysics Center (WIPAC) and Department of Physics,\\ University of Wisconsin, Madison, WI 53706, USA}

\author{Francis~Halzen} 
\affiliation{Wisconsin IceCube Particle Astrophysics Center (WIPAC) and Department of Physics,\\ University of Wisconsin, Madison, WI 53706, USA}

\begin{abstract}
The observed flux of ultra-high energy (UHE) cosmic rays (CRs) guarantees the presence of high-energy cosmogenic neutrinos that are produced via photo-hadronic interactions of CRs propagating through intergalactic space. This flux of neutrinos doesn't share the many uncertainties associated with the environment of the yet unknown CR sources. Cosmogenic neutrinos have nevertheless a strong model dependence associated with the chemical composition, source distribution or evolution and maximal injection energy of UHE CRs. We discuss a lower limit on the cosmogenic neutrino spectrum which depends on the observed UHE CR spectrum and composition and relates directly to experimentally observable and model-independent quantities. We show explicit limits for conservative assumptions about the source evolution.
\end{abstract}


\maketitle

\section{Introduction}

Cosmogenic neutrinos are produced when UHE CRs interact with the cosmic radiation background while propagating between their sources and Earth. The frequent interactions with the cosmic microwave background (CMB) limits the propagation of nucleons with energies greater than $E_{\rm GZK}\simeq40$~EeV to within a few 100~Mpc and is responsible for the so-called Greisen-Zatsepin-Kuzmin (GZK) cutoff of extra-galactic protons~\cite{Greisen:1966jv,Zatsepin:1966jv}. Mesons produced in these interactions quickly decay and produce an observable flux of cosmogenic (or GZK) neutrinos~\cite{Berezinsky:1970xj}. In fact, the observed spectrum of CRs extending up to energies of a few 100~EeV shows a suppression above $\sim E_{\rm GZK}$ with high statistical significance~\cite{Abbasi:2007sv,Abraham:2008ru}. This could be an indication that protons are dominating the flux at these energies. In this case the flux of cosmogenic neutrinos is typically large.

However, the experimental situation is less clear. Measurements of the elongation rate distribution of UHE CR showers indicate a transition of their arrival composition from light to heavy within 4-40~EeV~\cite{Abraham:2010yv,Unger:2011ry}. If a heavy component dominates also at higher energies the prospect for cosmogenic neutrino production is  ``disappointing''~\cite{Aloisio:2009sj} or at least less favorable than for the proton scenario. A crucial uncertainty of this scenario is the maximal injection energy of the nucleus with mass number $A$; as long as $E_{\rm max} \gg AE_{\rm GZK}$, even this scenario will produce an appreciable amount of cosmogenic neutrinos~\cite{Ahlers:2011sd}. If this condition is not met interactions with the subdominant cosmic photon background from the optical/infra-red will still contribute to the cosmogenic neutrino flux. We will use the estimate of Ref.~\cite{Franceschini:2008tp} for our calculation.

The IceCube neutrino observatory has reached the sensitivity for the detection of optimistic cosmogenic neutrino fluxes~\cite{Abbasi:2011ji}. In the case of a non-observation it is of interest to know a lower limit on the various source emission possibilities for their definite exclusion. Lower cosmogenic neutrino flux limits have already been discussed in the context of proton-dominated scenarios via a deconvolution of early Auger data~\cite{Fodor:2003ph}. We will discuss in this article updates of these lower limits and extensions to more general assumptions for the source distribution and chemical composition. Similar to Ref.~\cite{Fodor:2003ph} we will not attempt to construct a specific source emission model that fits the Auger spectrum and elongation rate distribution but we will derive the limits directly from the observed composition measurement and spectrum. From this we can derive a strict lower limit on the cosmogenic flux. 

\section{Cosmic Ray Propagation}

The propagation of UHE CR nuclei is affected by photo-disintegration~\cite{Stecker:1969fw,Puget:1976nz}, photo-hadronic interactions\cite{Mucke:1999yb}, Bethe-Heitler pair production~\cite{Blumenthal:1970nn} and red-shift losses due to the expansion of the Universe. It is convenient to consider a homogenous and isotropic distribution of CR sources and derive the observed CR from the co-moving number density $Y_i \equiv n_i/(1+z)^3$ as a solution to a set of Boltzmann equations~\cite{Ahlers:2009rf},
\begin{multline}\label{eq:diff0}
\dot Y_i = \partial_E(HEY_i) + \partial_E(b_iY_i)\\-\Gamma^{\rm tot}_{i}\,Y_i
+\sum_j\int{\rm d} E_j\,\gamma_{ji}Y_j+\mathcal{L}_i\,.
\end{multline}
The cosmic expansion rate $H(z)$ follows the usual ``concordance model'' dominated by a cosmological constant with $\Omega_{\Lambda} \sim 0.73$ and a (cold) matter component, $\Omega_{\rm m} \sim 0.27$ with $H^2 (z) = H^2_0\,[\Omega_{\rm m}(1 + z)^3 + \Omega_{\Lambda}]$, normalized to its value today of $H_0 \sim72$ km\,s$^{-1}$\,Mpc$^{-1}$~\cite{Nakamura:2010zzi}. The first and second terms on the r.h.s.~of Eq.~(\ref{eq:diff0}) describe, respectively, red-shift and other continuous energy losses (CEL) with rate $b \equiv -\mathrm{d}E/\mathrm{d}t$. In the following we will treat Bethe-Heitler pair production as a CEL process~\cite{Blumenthal:1970nn}. The third and fourth terms describe more general interactions involving particle losses ($i \to$ anything) with total interaction rate $\Gamma^{\rm tot}_i$, and particle generation of the form $j\to i$ with differential interaction rate $\gamma_{ij}$.  The last term on the r.h.s., $\mathcal{L}_i$, corresponds to the emission rate density of CRs of type $i$ per co-moving volume. The detailed description of the interaction rates and their scaling with red-shift has been discussed in our previous publications~\cite{Ahlers:2009rf,Ahlers:2010ty}. 

We first discuss the case of proton sources. The flux of cosmogenic neutrinos today ($z=0$) depends on the co-moving number density of protons at all red-shifts and can be approximated as~\cite{Ahlers:2009rf}
\begin{multline}\label{eq:Jnu}
J_\nu(E_\nu) \simeq\frac{1}{4\pi}\int_0^{\infty}
\frac{{\rm d} z'}{H(z')}\\\times\int\mathrm{d} \mathcal{E}_p\,\gamma_{p\nu}(z',\mathcal{E}_p,(1+z')E_\nu)\,Y_p(z',\mathcal{E}_p)\,,
\end{multline}
where $\mathcal{E}_p$ is the solution to the differential equation $\dot{\mathcal{E}}_p=-H\mathcal{E}_p -b_{\rm BH}(z,\mathcal{E}_p)$ with initial condition $\mathcal{E}_p(0,E_p) = E_p$. The co-moving number density of protons can be written as
\begin{multline}\label{eq:Yp}
Y_p(z,\mathcal{E}_p(z)) \simeq\frac{1}{1+z}\int_z^{\infty}\frac{{\rm d} z'}{H(z')}\mathcal{L}_{p,{\rm eff}}(z',\mathcal{E}_p(z'))\\\times\exp\left[\int_z^{z'}{\rm d} z'' \frac{\partial_Eb_{\rm BH}(z'',\mathcal{E}_p(z''))-\Gamma(z'',\mathcal{E}_p(z''))}{(1+z'')H(z'')}\right]\,,
\end{multline}
where the effective source term is defined as
\begin{multline}
  \mathcal{L}_{p, {\rm eff}}(z,E_p)\\ =\mathcal{L}_p(z,E_p) +\int\mathrm{d}
  \mathcal{E}_p\,\gamma_{pp}(z,\mathcal{E}_p,E_p)Y_p(z,\mathcal{E}_p)\,.\label{eq:Leff}
\end{multline}

\section{Minimal Neutrinos from Protons}

\begin{figure}[t]\centering
\includegraphics[width=\linewidth]{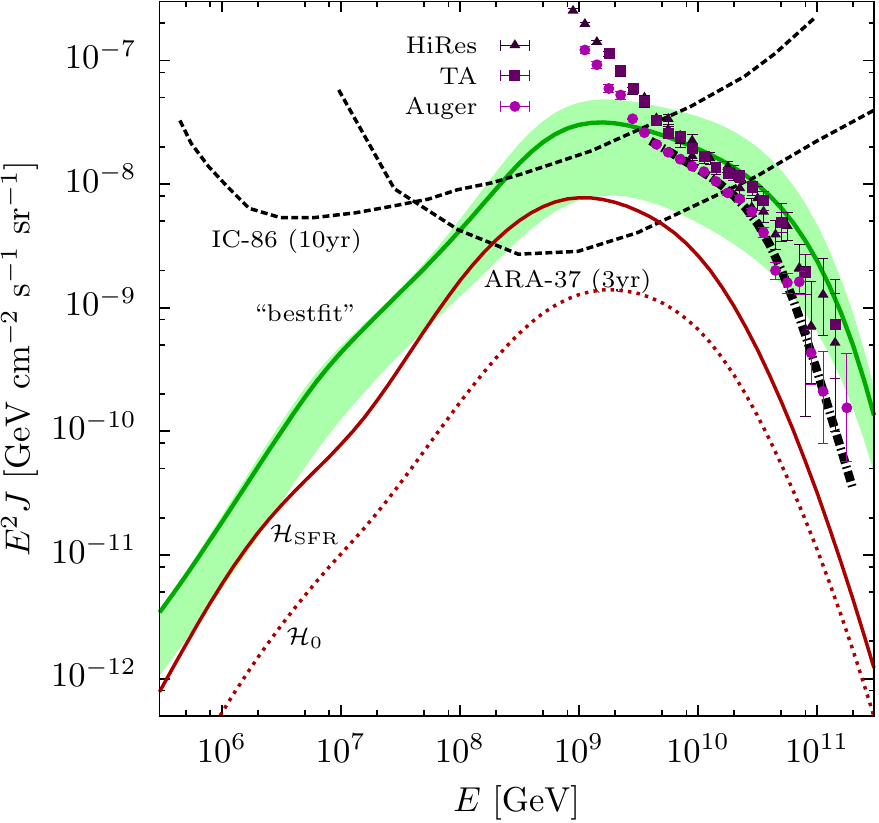}
\caption[]{Minimal flux of cosmogenic neutrinos assuming dominance of protons above 4~EeV. We show the results without source evolution (dotted) and assuming source evolution according to the star formation rate (solid). Also shown are the projected sensitivities of IceCube (10 years) and the ARA-37 (3 years) as dashed lines. The thick dashed-dotted line shows the approximation of the Auger spectrum above the ankle. For comparison, we also show the bestfit cosmogenic neutrino flux (green solid line) from Ref.~\cite{Ahlers:2010fw} ($E_{\rm min} = 10^{18.5}$~eV) including the 99\% C.L.~(green shaded area) obtained by a fit to the HiRes spectrum.}\label{fig1}
\end{figure}

A minimal contribution to the flux of cosmogenic neutrinos can be estimated as follows. As a first step we approximate the UHE CR spectrum measured by Auger via the phenomenological fit given in Ref.~\cite{Abraham:2010mj}. This fit is shown in Fig.~\ref{fig1} as a dashed-dotted line together with recent data of Auger, HiRes~\cite{Abbasi:2007sv} and the Telescope Array~\cite{AbuZayyad:2012ru} (TA). Note, that the normalization of the Auger data is lower by a about a factor two than HiRes and TA and hence cosmogenic neutrinos derived from this data are the lowest. 

Whereas the spectrum of UHE CRs is dominated by closeby sources, the neutrino flux receives contributions up to the Hubble scale. The overall flux will hence increase for an increasing number of sources with red-shift. We assume that redshift evolution decouples from the source emission spectrum, {\it i.e.} $\mathcal{L}_p(z,E) = \mathcal{H}(z)Q_p(E)$ and we consider two scenarios for the source evolution $\mathcal{H}(z)$. In the most conservative case we assume source contributions within redshift $z_{\rm max}=2$ with no source evolution, {\it i.e.}~$\mathcal{H}_0=\Theta(z_{\rm max}-z)$. A more realistic scenario assumes a source evolution following the star formation rate. We will use the estimate~\cite{Hopkins:2006bw,Yuksel:2008cu}
\begin{equation}\label{eq:HSFR}
\mathcal{H}_{\rm SFR}(z) = \begin{cases}(1+z)^{3.4}&z<1\,,\\
N_1\,(1+z)^{-0.3}&1<z<4\,,\\N_1\,N_4\,(1+z)^{-3.5}&z>4\,,
\end{cases}
\end{equation}
with normalization factors, $N_1 = 2^{3.7}$ and $N_4 = 5^{3.2}$. 
Since we assume conservative choices of the source evolution the associated cosmogenic neutrino flux can be regarded as {\it lower limits} on the expected cosmogenic neutrino flux.

In the following we will derive approximate solutions to Eqs.~(\ref{eq:Jnu}) and (\ref{eq:Yp}) using an iterative scheme. For the iteration start we choose $Q_p^{(0)}(E_p) = (H_0+\partial_Eb_0+\Gamma_0)4\pi J^{\rm obs}_{\rm CR}(E_p)$, where $b_0$ and $\Gamma_0$ are the energy loss and interaction rate, respectively, at redshift $z=0$.
The iteration step is then given by 
\begin{equation}
Q^{(n+1)}_p(E_p) =4\pi J^{\rm obs}_{\rm CR}(E_p)/\eta^{(n)}(E_p) \,,
\end{equation}
where we use the phenomenological fit of Ref.~\cite{Abraham:2010mj} for $J^{\rm obs}_{\rm CR}(E)$ and introduce the effective survival distance
\begin{multline}\label{eq:eta}
\eta^{(n)}(E_p) =\int_0^{\infty}\frac{{\rm d} z'}{H(z')}\frac{\mathcal{L}_{p, {\rm eff}}^{(n)}(z',\mathcal{E}_p(z'))}{Q_p^{(n)}(E_p)}\\\times\exp\left[\int_0^{z'}{\rm d} z'' \frac{\partial_Eb_{\rm BH}(z'',\mathcal{E}_p(z''))-\Gamma(z'',\mathcal{E}_p(z''))}{(1+z'')H(z'')}\right]
\,.
\end{multline}
We continue this iteration until the relative correction $\sum_i(Q_{{\rm p},i}^{(n+1)}/Q_{{\rm p}, i}^{(n)}-1)^2$ stops to decrease or a maximal (sufficiently large) iteration step is achieved. This compensates for numerical instabilities. 

In Figure~\ref{fig1} we show the resulting cosmogenic neutrino flux for this procedure for the two evolution scenarios. The limit for the SFR evolution agrees well with that derived from a deconvolution analysis in Ref.~\cite{Fodor:2003ph}. We also indicate in this plot the sensitivity of IceCube~\cite{Abbasi:2011ji} and the proposed Askaryan Radio Array (ARA)~\cite{Allison:2011wk}. Three years of observation with the 37 station configuration of ARA (``ARA-37'') is sufficient to reach the proton emission model for the SFR case. In the case of no source evolution this scenario is reached after ten years. As we already emphasized, this result depends on the absolute normalization and/or energy calibration of the observed UHE CR spectrum. For a normalization to HiRes and TA data we expect our limits to scale up by about a factor 2.

\section{Generalization to Heavy Nuclei}

The case of a more general scenario including UHE CR sources of heavy nuclei is more complicated. The chemical composition observed at Earth is the result of rapid photo-disintegration in the radiation background and there is no simple connection to the source composition. However, since photo-disintegration conserves the energy per nucleon we can derive a lower neutrino limit by tracking the leading (heaviest) nucleus back to its source starting from a composition $A_o$ and $Z_o$ inferred from UHE CR observations. 

The parent nuclei during this back-tracking are at least as heavy as the observed mass composition. For instance, a single helium nucleus in the observed spectrum might be produced via the production chain ${}^{10}{\rm B}\to{}^{9}{\rm Be}(+p)\to{}^{4}{\rm He}(+{}^4{\rm He}+p)$ from the source. The parent nuclei in each step of this chain determine the interaction and energy loss rates during propagation. For a lower limit on the cosmogenic neutrino flux we have to {\it minimize} the emission rate density of the UHE CR nuclei associated with their cascades in the CMB. This corresponds to a {\it maximal} survival probability of nucleons. Hence, we can derive a strict lower limit with the assumption that the back-tracking of the nuclei is indefinite, {\it i.e.}~we assume no upper limit on the atomic mass number in the nuclei cascades.

Photo-disintegration that drives the cascades competes with photo-hadronic interactions and Bethe-Heitler energy loss. To first order, a photo-hadronic interaction of the nucleon with energy $E$, charge $Z$ and mass number $A$ can be approximated via the interaction rate of the free proton as $\Gamma_{A\gamma}(E) \simeq A\Gamma_{p\gamma}(E/A)$. Hence, the interaction rate per nucleon of the parent nucleus is the same. Energy loss via Bethe-Heitler pair production, however, scales as $b_{A\gamma}(E) \simeq Z^2b_{p\gamma}(E/A)$ and the effective energy loss per nucleon scales as $Z^2/A$. Again, for a maximal survival probability of the nucleons and hence a minimal emission rate density of the sources, we assume a minimal Bethe-Heitler energy loss of the nucleons. This corresponds to the energy loss of a nucleus with charge $Z_o$ and atomic mass number $A_o$ associated with the observed composition.

\begin{figure*}[t]\centering
\includegraphics[width=0.48\linewidth]{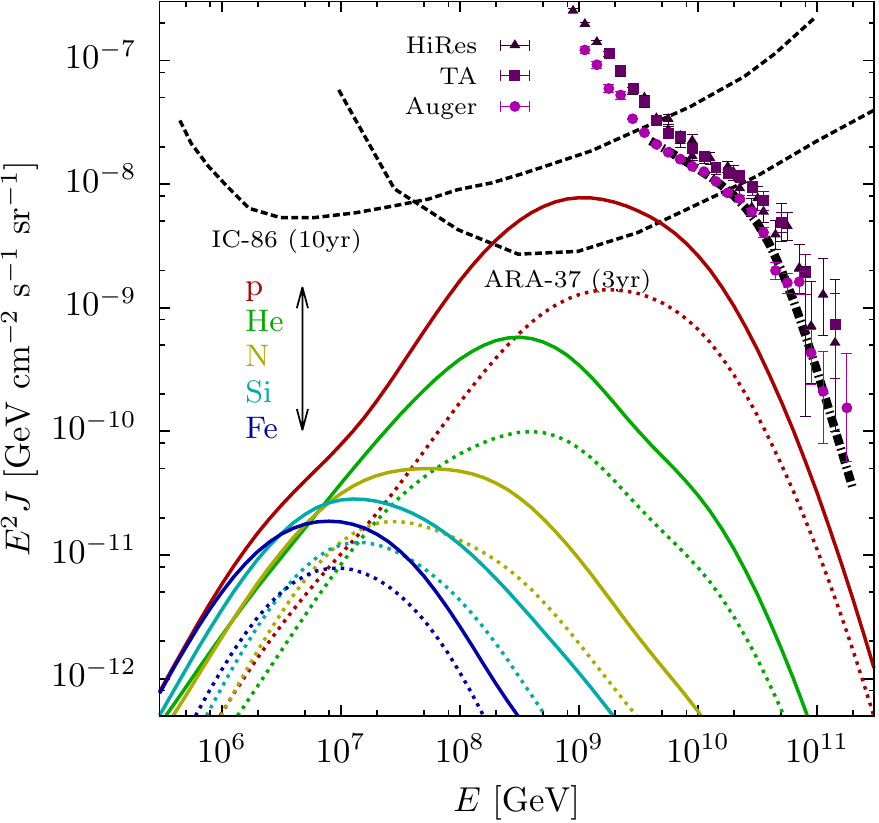}\hfill\includegraphics[width=0.48\linewidth]{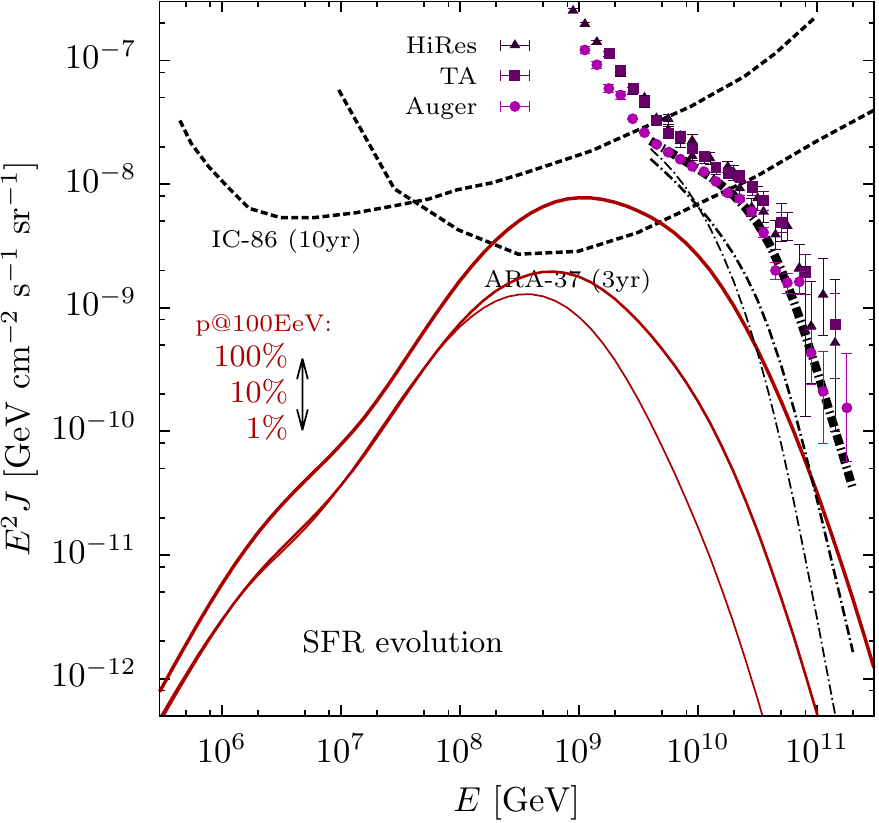}
\caption[]{Minimal flux of cosmogenic neutrinos for a mixed composition. {\bf Left panel:}  Minimal flux of cosmogenic neutrinos assuming dominance of protons, helium, nitrogen, silicon or iron in UHE CRs above 4~EeV. We show the results without source evolution (dotted) and assuming source evolution according to the star formation rate (solid). {\bf Right panel:} The contribution of protons (red lines) in a mixed composition scenario assuming 100\% (upper line), 10\% (middle line) and 1\% (lower line) proton contribution (black lines) at 100~EeV.}\label{fig2}
\end{figure*}

In summary, a lower limit on the cosmogenic neutrino flux can hence be derived by the same Eqs.~(\ref{eq:Jnu}) and (\ref{eq:Yp}) where we now replace the continuous energy loss by its minimal contribution $b_{\rm min}(z,E) \simeq (Z_o^2/A_o)b_{\rm BH}(z,E)$, where $b_{\rm BH}$ correspond to the energy loss of a free proton. The photohadronic interaction of the nucleons is given by the average interaction of protons and neutrons. The total number of nucleons per nucleon energy depends on the observed (or inferred) mass composition of UHE CRs. Assuming a single component we have the relation $E_NJ_N(E_N) = A_oE_{\rm CR}J_{\rm CR}(E_{\rm CR})$ with $E_N = E_{\rm CR}/A_o$ or $J_N(E_N) = A_o^2J_{\rm CR}(E_{\rm CR})$. 

In the left panel of Fig.~\ref{fig2} we show the minimal cosmogenic neutrino fluxes for the case of helium, nitrogen, silicon and iron dominance of the Auger spectrum. The level of these fluxes is not in reach of present or future neutrino observatories. However, cosmogenic neutrino fluxes strongly depend on the maximal injection energy of the sources. We conservatively assume for our method that the maximal energy does not exceed the observed energy of UHE CRs. However, it is in principle possible that these models produce detectable fluxes of cosmogenic neutrinos~\cite{Ahlers:2011sd} if the maximal energy significantly exceeds $A\times E_{\rm GZK}$. We will briefly discuss this in the following section.

We can also generalize our method to the case of a mixed compositions, which is indicated by the Auger CR elongation rate distribution. For instance, if $f_i(E_{\rm CR})$ denotes the fraction of nuclei with mass $A_i$ at CR energies $E_{\rm CR}$ the mean mass number is given by
\begin{equation}
J_N(E_N) \simeq \sum A_i^2 f_i(A_iE_N)J_{\rm CR}(A_iE_{\rm N})\,.
\end{equation}
Hence the minimal cosmogenic neutrino flux in this case is $J^{\rm min}_\nu(E_\nu) = \sum_i J^{\rm min}_i(E_\nu)$, where the individual $J^{\rm min}_i$ are derived in the same way as before but using $f_i(E_{\rm CR})J_{\rm CR}(E_{\rm CR})$ as the input spectrum. 
As an example we show in the right panel of Fig.~\ref{fig2} the lower limit associated with protons in a multi-component model, where we decrease the proton contribution at 100~EeV to 10\% ($\alpha=1$) and 1\% ($\alpha=2$) using $f_p = 1-(1+(E/10^{19}{\rm eV})^{-\alpha})^{-1}$ with $f_{\rm A} = 1-f_p$. 

\section{Optimistic Cosmogenic Neutrinos}

Predictions of the cosmogenic neutrino spectra are very sensitive to the maximal energy of UHE CR nuclei. In the following we will briefly discuss ``optimistic'' predictions that assume that the maximal energy of CR nucleons is much larger than the GZK cutoff, {\it i.e.}~$E_{\rm CR}/A\gg E_{\rm GZK}$. 
For the discussion it is convenient to introduce the energy density (eV cm${}^{-3}$) of the GZK neutrino background at redshift $z$ defined as
\begin{equation}
\label{eq:omegaGZK}
\omega_{\rm GZK}  \equiv \int {\rm d}E_\nu E_\nu Y_\nu(E_\nu)\,.
\end{equation}
From the Boltzmann equations~(\ref{eq:diff0}) we can derive the evolution of the energy density as
\begin{equation}\label{eq:omegaevol}
\dot\omega_{\rm GZK} + H\omega_{\rm GZK} = \sum_i\int{\rm d} E\, b_{i, {\rm GZK}}(z,E)Y_i(z,E)\,,
\end{equation}
where $b_{i,{\rm GZK}}(E)\simeq 0.2 E \Gamma_{\gamma\pi}(E/A_i)$ is an approximation of the energy loss of the nuclei into GZK neutrinos~\cite{Ahlers:2011sd}.

The UHE CR interactions with background photons are rapid compared to cosmic time-scales. The energy threshold of these processes scale with redshift $z$ as $A_iE_{\rm th}/(1+z)$ where $E_{\rm th}\gtrsim E_{\rm GZK}$ is the (effective) threshold today. We can therefore approximate the evolution of the energy density as
\begin{equation}
\dot\omega_{\rm GZK} + H\omega_{\rm GZK} \sim \frac{3K_\pi\mathcal{H}(z)}{4(1+K_\pi)}{\sum}_i\!\!\!\!\!\!\!\!\!\!\int\limits_{A_iE_{\rm th}/(1+z)}\!\!\!\!\!\!\!\!\!\!{\rm d} E\,E\,Q_i(E)\,,
\end{equation}
where $K_\pi$ is the ratio of charged to neutral pions produced in $p\gamma$ interactions. Assuming a power-law emission rate density $Q_i(E)\propto E^{-\gamma_i}$ with sufficiently large cutoff $E_{\rm max}\gg E_{\rm th}$ we see that cosmic evolution enhances the GZK flux as
\begin{equation}\label{eq:evolfactor2}
\omega_{\rm GZK} \sim \frac{3}{8} \sum_i\eta_i\frac{(A_iE_{\rm th})^2Q_i(A_iE_{\rm th})}{\gamma_i-2}\,,
\end{equation}
where the last term assumes $\gamma_i>2$ and the effective survival distance of the nucleons is defined as
\begin{equation}
\eta_i = \int_{0}^{\infty}\frac{{\rm d}z}{H(z)}\mathcal{H}(z)(1+z)^{\gamma_i-4}\,.
\end{equation}
For $\gamma_i\simeq2$ and for those evolution scenarios $\mathcal{H}$ that we have considered so far in this paper, the effective survival distances range from $0.48/H_0$ (no evolution) to $2.4/H_0$ (SFR). This agrees well with the relative ratio $\sim5$ of the energy densities associated with lower neutrino limits in the proton-dominated scenario shown in Fig.~\ref{fig1}.

The relation~(\ref{eq:evolfactor2}) shows that as long as the maximal energy per nucleon is much larger than the pion production threshold in the CMB ({\it i.e.}~$E_{\rm max} \gg AE_{\rm GZK}$) and the injection index is $\gamma_i\simeq2$ the main difference in the energy density of GZK neutrinos comes from the underlying evolution model, not by the inclusion of heavy elements. In principle, this factor can be large even for heavy nuclei if the sources have a strong evolution. The fact that typical CR models including heavy nuclei produce significantly less GZK neutrinos can be traced back to a low maximal energy per nucleon and/or a weak evolution of CR sources. Note that the latter is an important ingredient of proton-dominated low-crossover models \cite{Berezinsky:2002nc}, whereas CR models of heavy nuclei including more model degrees of freedom are less predictable w.r.t.~the source evolution.

Note that, ultimately, the inferred energy density $\omega_{\rm \gamma}$ of the extragalactic diffuse $\gamma$-ray background in the GeV-TeV region constitutes an upper limit for the total electro-magnetic energy from pion-production of UHE CR nuclei, see {\it e.g.}~\cite{Ahlers:2010fw}. An upper limit is given via the relation
\begin{equation}\label{eq:cas1}
\omega_{\gamma} \gtrsim \left(\frac{1}{3}+\frac{4}{3K_\pi}\right)\omega_{\rm GZK}\,.
\end{equation}
Recent result from Fermi-LAT~\cite{Abdo:2010nz} translates into an energy density of $\omega_{\gamma} \simeq 6\times10^{-7} {\rm eV}/{\rm cm}^{3}$ \cite{Berezinsky:2010xa,Ahlers:2010fw}. Assuming an $E^{-2}$ neutrino spectrum between energies $E_-$ and $E_+$ a numerical simulation gives a {\it cascade limit} of
\begin{equation}\label{cascade}
  E^2J^{\rm cas}_{{\rm all}\,\nu}(E) \simeq \frac{3\times10^{-7}}{\log_{\rm 10}(E_+/E_-)}\,{\rm GeV}\,{\rm cm}^{-2}\,{\rm s}^{-1}\,{\rm sr}^{\,-1}\,\,.
\end{equation}
This is only slightly lower than the estimate~(\ref{eq:cas1}) for $K_\pi=1$. Cosmogenic neutrino fluxes that saturate this bound in the EeV region are already ruled out by IceCube upper limits~\cite{Abbasi:2011ji}.

\section{Discussion}

We have discussed in this article a simple procedure to derive lower limits on the cosmogenic neutrino flux. The limits are based on the observed spectrum and composition of UHE CRs and depend on the unknown evolution of sources. For the case of a proton-dominance in the UHE CR data we show that ARA-37 should identify the flux of cosmogenic neutrinos after 3 years of observation if UHE CR sources follow the star formation rate. For the less optimistic (and less realistic) case of no source evolution it would require 10 years of observation.

In the case of heavy nucleus dominance of the CR flux cosmogenic neutrino predictions are less optimistic. We can derive a lower limit in this scenario by tracking the leading nucleus back to its source. Since photo-disintegration conserves the energy per nucleon of the interaction we can base our analysis on the observed number of nucleons in UHE CRs, which depends on the observed mass composition.

The dominant contribution to the cosmogenic neutrino flux is expected from the proton content in the UHE CR spectrum. We show in Fig.~\ref{fig2} two cases where we decrease the contribution of protons to 10\% and 1\% at 100~EeV and assume source evolution with the star-formation rate. Even this less optimistic case is in reach of ARA-37 after 5 years of observation.

The prediction of cosmogenic neutrinos is very sensitive to the maximal CR injection energy per nucleon. If this is significantly larger than the GZK cutoff, even UHE CR scenarios dominated by heavy nuclei can produce large fluxes of cosmogenic neutrinos. For flat spectra that are sufficiently close to $E^{-2}$ the energy density of these optimistic GZK neutrino predictions depends on the cosmic evolution of the sources.

All cosmogenic neutrino fluxes shown in this analysis are normalized to Auger data. The spectra observed with HiRes and the Telescope are in general larger, which could be a result of an overall systematic energy shift by $20-30$\%. This corresponds to an upward shift of up to a factor 2 of the energy density $E_{\rm CR}^2J_{\rm CR}(E_{\rm CR})$. Hence the lower limits shown in Figs.~\ref{fig1} and \ref{fig2} should be similarly scaled upward. 

Finally, we would like to stress that the present analysis does not take into account statistical uncertainties of the CR data. However, the method can be easily extended to this case. In Ref.~\cite{Ahlers:2010fw} we have shown that an actual fit to HiRes data assuming a proton power-law injection in the sources is statistically consistent with cosmogenic neutrino fluxes that exceed the minimal bound by up to an order of magnitude and are in reach of the IceCube detector.

\noindent {\it Acknowledgments.} We thank Albrecht Karle, Ali Kheirandish and Kohta Murase for discussion. MA acknowledges support by a John Bahcall Fellowship for neutrino astronomy of the Wisconsin IceCube Particle Astrophysics Center (WIPAC). FH is supported in part by the National Science Foundation under Grant No.~OPP-0236449, by the DOE under grant DE-FG02-95ER40896 and in part by the University of Wisconsin Alumni Research Foundation.

\end{document}